\documentstyle[psfig,pre,aps]{revtex}
\begin{document}

\def\fig#1{}
\def\pacs#1{PACS: #1}
\def\address#1{}

%\draft

\author{
Zolt\'an Csah\'ok and Tam\'as Vicsek
} 

\title{Lattice-gas model for collective biological motion} 

\address{Department of Atomic Physics, E\"otv\"os University,\\
Budapest, Puskin u. 5-7, HUNGARY H-1088
}

\maketitle

\begin{abstract}
A simple self-driven
lattice-gas model for collective biological motion is introduced.
We find weakly first order phase transition from individual
random walks to collective migration.
A mean-field theory is presented to support the numerical results.
\end{abstract}

\pacs{05.50.+q, 64.70.-p, 05.60.+w}
%%%
%%%	64.70	Phase transitions
%%%	05.50	Lattice theory and statistics
%%%	05.60+w	Transport processes: theory

\section{Introduction}
\label{sec_intro}

One of the most interesting aspects of evolution is the emergence of
multicellular organisms due to the appearance of cooperation and
differentiation of eucariotes.  Although much research has been done
along this line,
some of the related basic questions are still open.  As a natural step
towards the understanding of the physical and physico-chemical
background of self-organization of microorganisms several authors
considered relatively simple systems such as the development of
bacterial colonies.

The growth of bacterial colonies having complex geometries
has been  extensively
	studied recently \cite{Matsu90,Ber91,Matsu93,Ben94,Med94}.
One of the mean aspects was the fractality \cite{Vic92}
of the growing colonies.
It has been  found that the framework of diffusion limited growth
fits well these phenomena \cite{Matsu90,Matsu93}.
In addition, the various morphologies  of growing colonies have been
experimentally investigated  recently \cite{Ben94}
and a dynamical
model has been introduced which incorporates a wide range
of effects
relevant to the phenomenon of collective bacterial growth and motion,
for example chemotaxis. In a further related model
 \cite{Vics95},
 aimed at describing the collective
motion of self-driven particles (such as bacteria),
a quasi-continuous approximation has been  used with rules
(particles moving with the same absolute velocity
take on the average
direction of motion of the neighboring particles)  based on
biological assumptions.
 As a main result it has been shown
that spontaneous breaking of rotational symmetry can occur
as the density of particles is increased or
the level of random noise (i.e., the temperature)
is lowered. The transition has been found
to be continuous.

One of the basic differences between  living and azoic systems is
that living organisms are {\it self-driven}: they can transform
energy gained from  food into mechanical energy which allows them
to change their position.
As the simplest example we can take
bacteria \cite{Gun60,Led92},
having various ways of motion.
One of the mechanisms is  motion
by the means of flagella: the bacteria
have flagella  functionally analogous to
a propeller attached to a motor.
The motion of organisms
is not under control of an external
field, as is common in physical systems. Instead,
  the environmental effects cause only
a change in the local velocity of the organisms.
Since living objects are capable of communicating
in various ways (ranging from
the sensing of chemicals  to verbal communication among humans),
 an organism is in continuous interaction not only with
its environment but also with other organisms in its neighborhood.
Thus, in the first approximation a system of
organisms can be considered as  an
open interacting multi-particle physical system.
Then, one can attempt to apply the methods recently developed in the
investigations of 
complex systems \cite{cplx88a,cplx88b}.

In this paper we present a simple lattice gas
model for the collective motion of self-driven particles.
Similar approach has been applied
to traffic systems \cite{Nag93,Csa94,Sch95}
which also belong to the class of self-driven systems.
Further approaches to self-driven systems include
reaction-diffusion description \cite{Lev91},
investigation of the related integrodifferential equations 
\cite{Ede90},
molecular dynamics \cite{Dup95}			%%  Deu93 ???
and cellular automata \cite{Erm91,Hem95}.

The aim of this paper is to extend
the usual statistical physical description for
a particular case of 
collective motion in systems of living objects.
First we introduce our model, then we give theoretical
description of the problem.
In Section\ \ref{sec_numres} we present the numerical results 
and in Section\ \ref{sec_concl} we summarize our results.

\section{The model}
\label{sec_model}

Our model is defined on triangular lattice of $L^2$ sites
with unit lattice spacing and periodic boundary conditions.
We put $N$ particles (bacteria) on the lattice, where $N$ is not
necessarily smaller than the number of lattice sites.
The density of the particles is defined as
\begin{equation}\label{rh}
\varrho = {N\over L^2}.
\end{equation}

Each site can be either empty or occupied by one or more particles.
If more than one particle is present at
a site then in the calculations only the lowest one will be
considered, where the lowest particle is defined
as having the smallest random number previously
assigned to each particle.

The particles are characterized by their position
${\bf r}_i$ and velocity ${\bf v}_i$ ($i=1\dots N$)
which is of unit length
($\vert {\bf v}_i \vert =1$)
and can point in any of the lattice directions
(${\bf u}_\alpha$, Fig.\ \ref{fig0}.).

At one time step positions and velocities of all particles 
are updated simultaneously according to the following rules:

\begin{enumerate}
\item
\begin{enumerate}
	\item for the particles which are not the lowest at
their position we assign a random direction;

\item for the particles which are the lowest at
their site we choose a new velocity  ${\bf u}_\alpha$
from a Boltzmann distribution;

$$ P({\bf u}_\alpha) = {1\over {\cal Z}}
{ \exp ( -\beta {\bf u}_\alpha \sum_{j\in {\rm lnn}} {\bf v}_j )},
$$
where ${\cal Z}$ is a normalizing factor so that
$\sum_{\alpha=1}^6 P({\bf u}_\alpha) = 1 $,
and $\beta$ is $1/T$ ($k_B=1$).
The summation goes over the nearest neighbors
which are in lowest position (lnn).
\end{enumerate}

\item every particle is moved one lattice unit in
direction of its velocity:
$$ {\bf r}_i \gets {\bf r}_i + {\bf v}_i. $$

\end{enumerate}

Note that the last step may result in sites
with occupancy higher than one,
this is the reason why we have to deal with such cases.
The motivation for Step 1(a) is that
we try to minimize the effect of multiple occupancy by letting
the extra particles to diffuse away.
The temperature parameter is not connected to the 
ambient
temperature of the bacterial colony, it is rather an effective
value which depends on many external parameters, as for example
food concentration and agar humidity. The case of high
food concentration is likely to be represented by high $T$
values since then the bacteria can move faster and do not
need coordinated behavior to extract food from the agar.
On the other hand, when there is a food shortage the bacteria
tend to cooperate, which results in a lower effective temperature. The
above model is in its spirit close
to the continuum model of self-driven particles \cite{Vics95},
however, the present version has a number of
new features which had to be introduced
because of its discrete nature.

One of the quantities of  interest is the average velocity
of the particles which we shall consider as the order parameter and
define as
\begin{equation}\label{m}
m = {1\over N} \left| \sum_i {\bf v}_i \right|.
\end{equation}
Obviously $0\le m \le 1$ holds.
To have a closer analogy with spin systems
we define a Hamiltonian
\begin{equation}\label{H}
H =  - {1\over 2} \sum_{i,j\in {\rm lnn}} {\bf v}_i {\bf v}_j,
\end{equation}
in accord with the simulation rule Step 1.
Those particles which are not the lowest at
their position do not give contribution
to the energy, they are regarded as a free gas.
The  energy per particle is
$$
\langle\varepsilon\rangle = {E\over N}.
$$
Having introduced the energy it is straightforward to define
the  heat capacity per particle
$$
c = {\partial \varepsilon \over \partial T}.
$$
Although we have similarities with spin systems our
model differs in a very specific way: the spins 
in our model 
are
{\it moving} and this spatial dynamics is
coupled to the spin dynamics.

Fig.\ \ref{fig1}. shows a possible time evolution of the position and
velocities of five
particles. The particles are lettered from {\tt A} to {\tt E} and
the arrows show the direction of their velocity (${\bf v}_i$).
At time step {\it b)} they form a cluster (containing a doubly
occupied site) which
then gradually breaks up.

\section{Theory}
\label{sec_theory}

Our system in closely related
to the 6-state Potts model \cite{Pot52} since
we have $q=6$ possible velocity states for a particle.
Unlike the Potts model these states are not orthogonal,
and we have an essentially non-equilibrium system,
nevertheless, a mean field theory can be constructed
in a similar way \cite{Wu92}.

First we introduce a mean field Hamiltonian  instead of
Eq.\ (\ref{H})

\begin{equation}\label{Hmf}
H_{\rm MF} =  - {1\over 2} \sum_{i,j\in{\rm l}} {\bf v}_i {\bf v}_j,
\end{equation}
where the summation goes over all {\it lowest} (l) particles, not
only the nearest neighbors.
The mean field energy function can be written as

\begin{equation}\label{Emf}
E_{\rm MF} = - {1\over 2} N \varrho_{\rm eff}
\sum_{\alpha,\gamma=1}^6 x_\alpha U_{\alpha\gamma}  x_\gamma,
\end{equation}
where $\varrho_{\rm eff}= 1-\exp(-\varrho)$ is the effective density,
i.e., a site has on average $6 \varrho_{\rm eff}$ occupied
neighboring sites, $x_\alpha$ is the fraction
of particles travelling in the lattice direction $\alpha$
 ($\sum_{\alpha=1}^6 x_\alpha=1$)
and 
$U_{\alpha\gamma}= {\bf u}_\alpha {\bf u}_\gamma$ which for
the case of the Potts model would be simply
$U_{\alpha\gamma}= \delta_{\alpha\gamma}$.

The average energy per particle is
\begin{equation}\label{epsavg}
 \varepsilon_{\rm MF}  = {E_{\rm MF}\over N} =
 - {1\over 2} \varrho_{\rm eff}
\sum_{\alpha,\gamma=1}^6 x_\alpha U_{\alpha\gamma}  x_\gamma.
\end{equation}
The  entropy per particle is
$$
s_{\rm MF} = - \sum_{\alpha=1}^6 x_\alpha \ln x_\alpha,
$$
so for the free energy per particle one gets
\begin{equation}\label{fe}
\beta f_{\rm MF} = \beta {F_{\rm MF}\over N} =
\sum_{\alpha=1}^6 \left( x_\alpha \ln x_\alpha
- {1\over 2} \varrho_{\rm eff} \beta
x_\alpha \sum_{\gamma=1}^6 U_{\alpha\gamma}  x_\gamma
\right).
\end{equation}

We intend  to find the configuration ${x_\alpha}$
which minimizes the free energy $f_{\rm MF}$.
Since all the lattice directions are equivalent
we can look for a solution in the form of
\begin{equation}\label{x1}
x_1 = {1\over 6} + {5\over 6}m_{\rm MF}
\end{equation}
and
\begin{equation}\label{x2}
x_{\alpha>1} = {1\over 6} - {m_{\rm MF}\over 6},
\end{equation}
where $m_{\rm MF}$ is the mean field order parameter
which satisfies the relation
$$
m_{\rm MF}= \left| \sum_{\alpha=1}^6 x_\alpha {\bf u}_\alpha \right|
$$
according to Eq.\ (\ref{m}).
Substituting Eq.\ (\ref{x1}) and Eq.\ (\ref{x2})  to Eq.\ (\ref{fe})
after a bit of algebra one gets for
the mean field free energy
\begin{eqnarray}
\beta f_{\rm MF} = &&-{1\over 2}\varrho_{\rm eff}\beta~m_{\rm MF}^2 
- \log{1\over 6}+ {5(1-m_{\rm MF})\over 6}\log{1-m_{\rm MF}\over 6}
\nonumber\\
&&{} +{1+5 m_{\rm MF}\over 6}\log{1+5 m_{\rm MF}\over 6}.
\end{eqnarray}

For high temperatures ($T>T_c$) this function has its
minimum at $m_{\rm MF}=0$ which corresponds to
the disordered state of the system.
At the critical temperature, which can be derived
from $f_{\rm MF}$ and in our case it is
\begin{equation}\label{tc}
T_c \approx {\varrho_{\rm eff}\over 3.353},
\end{equation}
a  non-trivial minimum appears.
The phase transition, like in the 6-state
Potts model \cite{Bin81}, is first order.
The jump in the order parameter
in this approximation is given exactly by
$$
\Delta m_{\rm MF} = 0.8.
$$

\section{Numerical results}
\label{sec_numres}

We have studied our systems by Monte-Carlo simulations.
For initial configuration we chose
random distribution for the position and
velocity of the particles.
Typical configurations of the system
for various temperatures $T$ and particle
densities $\varrho$ are shown in Fig.\ \ref{fig2}.
It can be easily seen that at  low temperature
the particles tend to form clusters as it can be expected.

We have performed several long-time runs to
obtain the behavior of the
quantities defined in Section\ \ref{sec_model}.
as a function of $T$ and $\varrho$.
We used various system sizes ($L$) up to 40 for
high densities and up to 200 for low densities.
The limiting factor was the convergence time which
for the case of our largest system was in order of
$10^6$ sweeps of the system.
Fig.\ \ref{fig3}. demonstrates the order parameter as a function
of the temperature for $\varrho=0.9$. The estimated jump at $T_c$
is smaller but close to the value obtained by the mean field
theory.
In Fig.\ \ref{fig4}. we present the average energy which is
also subject to a finite jump at $T_c$.
These two figures suggest that a first order phase transition
takes place at $T=T_c$ in agreement with the theoretical
prediction.
A strong evidence supporting this idea is presented
in Fig.\ \ref{fig5}.
where we have plotted the distribution $P(\varepsilon)$ of the energy
values for a number of different temperatures below and above
$T_c$.
One can clearly see a gap in the distributions at intermediate
energies which is
a unique feature of first order phase transitions \cite{Bin92}.
The inset in the figure shows the distribution for Ising type
interaction of non-moving particles in the same system
where the transition is known to be second order.
In Fig.\ \ref{fig6}. we present the temperature dependence of the heat
capacity which is the measure of the broadness of the energy
distribution. A characteristic peak can be observed near $T_c$.
The position of the peak is shifted for various lattice sizes
due to finite size effects.

We studied the behavior of the model also as a function of
density of particles ($\varrho$).
Fig.\ \ref{fig7}. shows the temperature dependence
of the average energy for various densities obtained.
The transition is present even for very small densities
although the position of the critical temperature lowers.
In Fig.\ \ref{fig8}. we have plotted the dependence of the
transition temperature on the density.
There is a natural distinction between the high and low
density regimes of the system:
at the percolation threshold the behavior of the system is
expected to change. In fact we observe a change in
the dependency of the critical temperature
below the percolation threshold of the triangular lattice
($\varrho= 1/2$ and $\varrho_{\rm eff}\approx 0.39$)
at $\varrho_{\rm eff}\approx 0.25$ which corresponds to
density $\varrho\approx  0.29$.
The values of the measured critical temperatures are higher
than the one obtained from Eq.\ (\ref{tc}) which shows
the boundaries of applicability of our mean-field approximation.

The behavior of the average energy of the Potts model
near its transition temperature 
can be characterized by two exponents
$\alpha^{(-)}$ and $\alpha^{(+)}$ \cite{Bin81}.
These exponents are present due to the weakly first
order nature of the transition.
The temperature dependence of the average energy
is given by 
\begin{equation}\label{aplus}
\langle\varepsilon\rangle =
\varepsilon^{(-)} - A^{(-)} ( 1 - T/T_c)^{1-\alpha^{(-)}}
\end{equation}
for $T<T_c$ and similarly
\begin{equation}\label{aminus}
\langle\varepsilon\rangle =
\varepsilon^{(+)} + A^{(+)} ( 1 - T_c/T)^{1-\alpha^{(+)}}
\end{equation}
for $T>T_c$.
The difference between
$\varepsilon^{(+)}$ and $\varepsilon^{(-)}$ is equal to the energy
jump during the phase transition.
In Fig.\ \ref{fig_a1}. and Fig.\ \ref{fig_a2} we plotted
the energy differences versus the temperature
according to Eq.\ (\ref{aplus}) and Eq. (\ref{aminus})
for $\varrho=0.9$.
The exponents obtained from the slopes
are
$$
\alpha^{(+)} \approx 1 - 0.73 = 0.27
$$
and
$$
\alpha^{(-)} \approx 1 - 0.5 = 0.5.
$$
These values are different both from 
those of the $q=6$ state Potts model
($\alpha^{(+)}\approx 0.7$ and $\alpha^{(-)}\approx 0.7$)
and in part different from the corresponding
mean field values 
($\alpha^{(+)}=\alpha^{(-)}= 1/2$).

\section{Conclusion}
\label{sec_concl}

We have presented a lattice-gas model for collective biological
motion.
We have shown both numerically and theoretically that
weakly first
order phase transition takes place in our system
separating the phase with zero net transport and
the ordered phase with non-zero average velocity.
We find the exponents $\alpha^{(+)}$ and $\alpha^{(-)}$ are
differ from the ones of the $q=6$ Potts model and
from mean field values.
This difference can be attributed to the fact that 
although we have similarities with spin systems
our model differs in a very specific way:
the spins in our model are {\it moving}
and this spatial dynamics is coupled
to the spin dynamics.
It is remarkable that the behavior of the present lattice model is
qualitatively different from that of the analogous continuum model 
\cite{Vics95}.
While in the continuum model and in a directly related
continuum equation for a two-dimensional dynamic XY model
\cite{Tu95} a second order transition
was observed, in our lattice gas version the transition is
more complex and has a first order component.  Such discrepancies,
however, are not unfamiliar even
in two-dimensional equilibrium systems:
in particular, there is no long range ordering
in the equilibrium XY model \cite{Kos73}
having continuous symmetry, while its discrete counterparts 
(i.e., the Ising model) exhibit second order phase transition.

 % NEW ________________

\section{Acknowledgements}
This research was supported by the Hungarian
Research Foundation Grant No. T4439  and by the EEC Human Capital
Mobility Programme through TEMPUS and the contract ERB-CHRX-CT92-0063.
One of the  authors (Z. Cs.) is grateful to R. Botet
for his kind hospitality during his visit at Universit\'e Paris-Sud IX.

 % NEW ^^^^^^^^^^^^^^^^

\eject

\section{Figures}

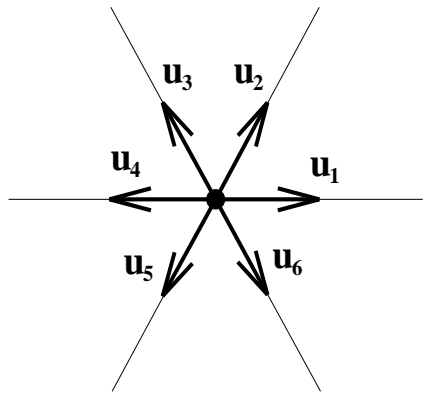
Fig.1. One lattice site, the six lattice directions are shown.

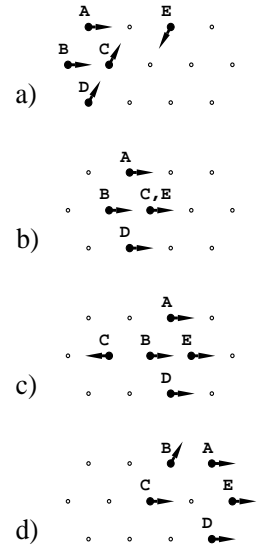
Fig.2. (a-d) Possible time evolution of our model ($a\to d$) shown
on a small portion of the lattice. Note the double occupancy in time step $b$.

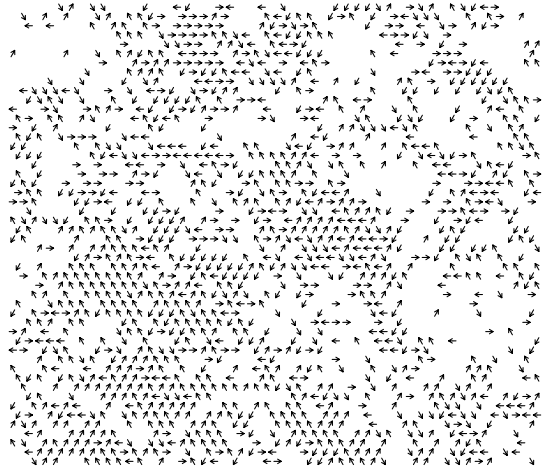
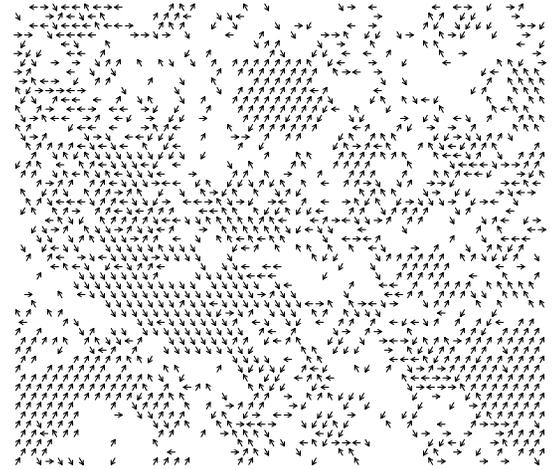
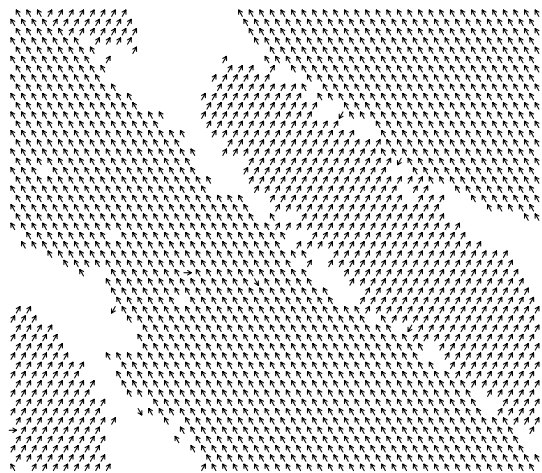
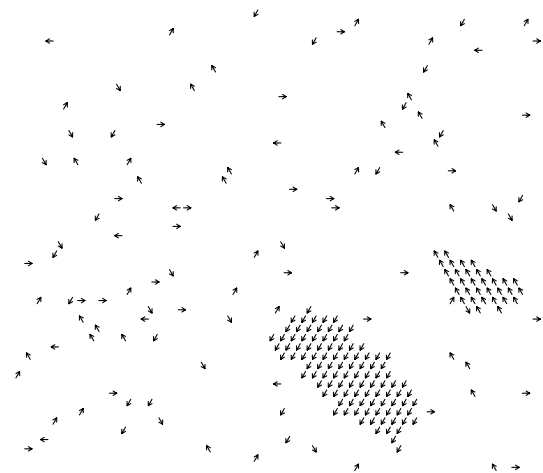
Fig. 3. (a-d) Some typical snapshots of the system  at (a) high temperature, 
(b) intermediate temperature, (c) low temperature at $\varrho= 0.9$ and 
(c) intermediate temperature at a lower density. Note the appearance of 
ordered clusters. (Only the particles in lowest position are drawn.)

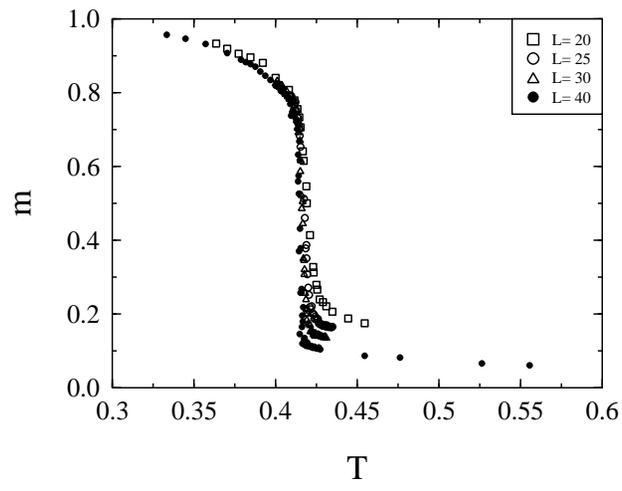
Fig.4. The order parameter ($m$) as a function of the temperature ($T$)
for density $\varrho=0.9$.

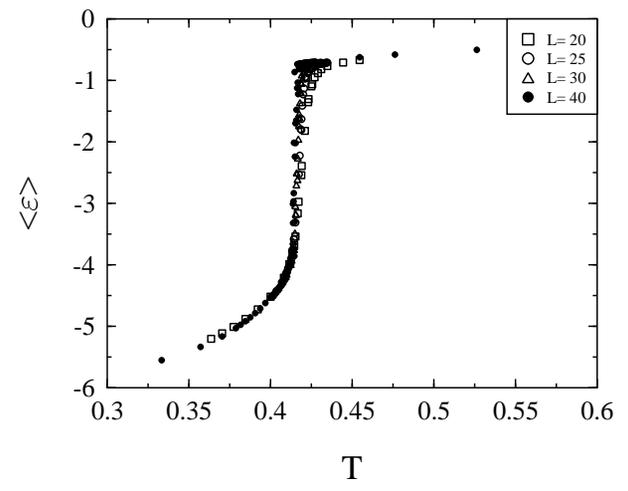
Fig.5. The average energy ($\langle\varepsilon\rangle$) versus
the temperature in the same systems as on the previous figure.

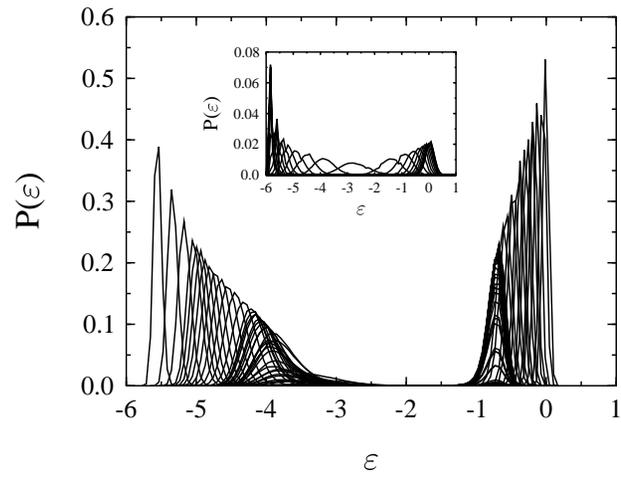
Fig.6. The energy distribution  ($P(\varepsilon)$) for various temperatures
below and above the transition ($\varrho=0.9$). Note the energy gap between
$\varepsilon\approx -1$ and $\varepsilon\approx -3.5$.
Inset shows the distribution for Ising spins instead of mobile
particles where the transition is continuous.

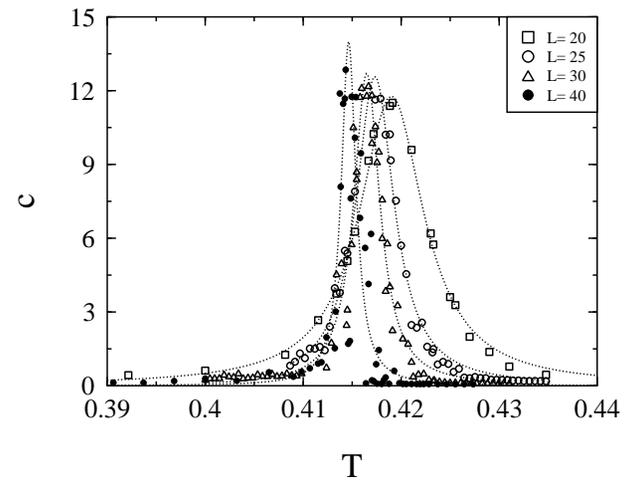
Fig.7. The heat capacity ($c$) versus temperature graph
for the systems as on Fig.4.
The dotted lines are guide to the eye.

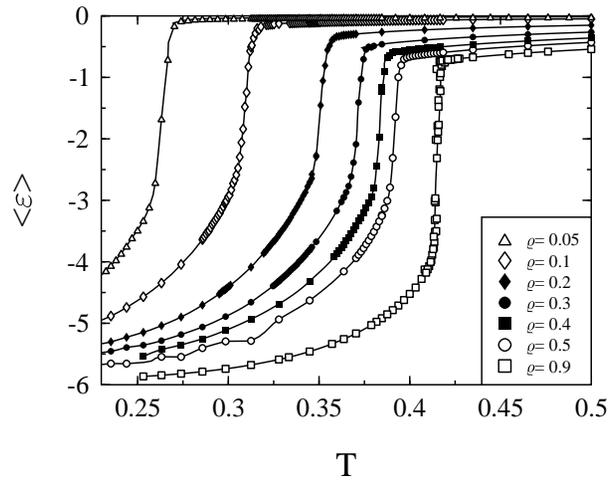
Fig.8. The average energy as a function of temperature for
various densities of particles.

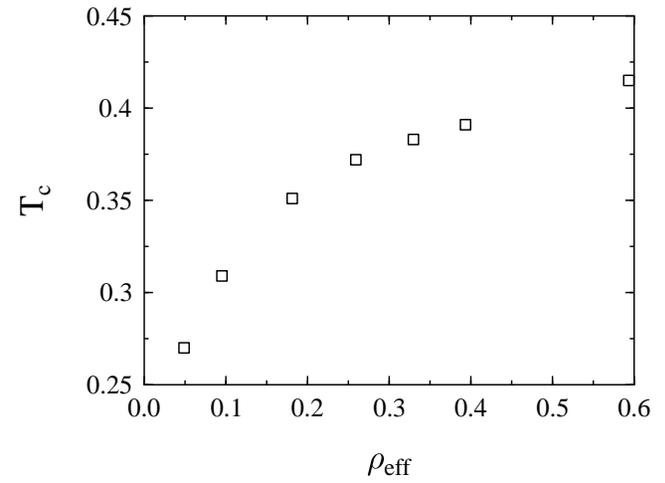
Fig.9. The critical temperature as a function of the
density of particles.

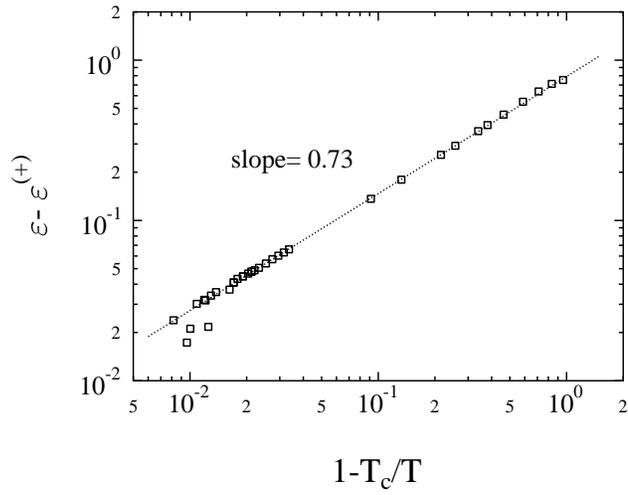
Fig.10. Energy difference versus temperature for $\varrho=0.9$.
($T_c= 0.413, \varepsilon^{(+)}= -0.76$)

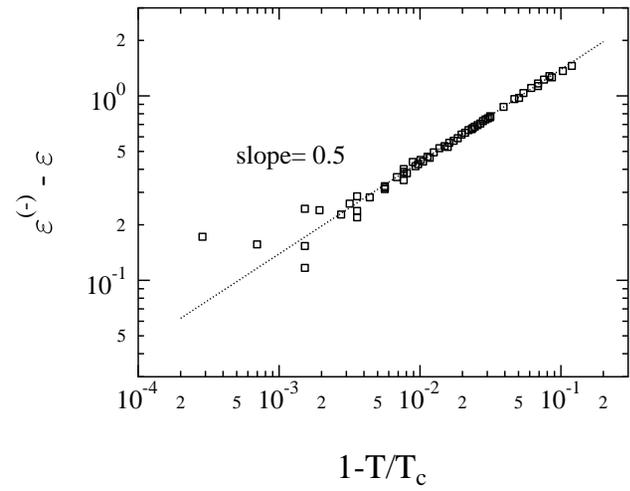
Fig.11. Energy difference versus temperature for $\varrho=0.9$.
($T_c= 0.413, \varepsilon^{(-)}= -3.75$)

\eject

\begin{figure}
\psfig{figure=c_fig.ps_page_1,width=15cm,height=20cm}
\end{figure} 
\begin{figure}
\psfig{figure=c_fig.ps_page_2,width=15cm}
\end{figure} 
\begin{figure}
\psfig{figure=c_fig.ps_page_3,width=15cm}
\end{figure} 
\begin{figure}
\psfig{figure=c_fig.ps_page_4,width=15cm}
\end{figure} 
\begin{figure}
\psfig{figure=c_fig.ps_page_5,width=15cm}
\end{figure} 
\begin{figure}
\psfig{figure=c_fig.ps_page_6,width=15cm}
\end{figure} 
\begin{figure}
\psfig{figure=c_fig.ps_page_7,width=15cm}
\end{figure} 

\end{document}